\documentclass{iau}
\usepackage{graphicx}

\title[] 
{HD141569A: disk dissipation caught in action}

\author[]   
{J.~P\'ericaud$^{1,2}$,
E.~Di Folco$^{1,2}$,
A.~Dutrey$^{1,2}$,
J.-C.~Augereau$^{3,4}$,
V.~Pi\'etu$^5$
 \and S.Guilloteau$^{1,2}$
 }

\affiliation{$^1$Univ. Bordeaux, LAB, UMR 5804, F-33270, Floirac, France\\email: {\tt jessica.pericaud@obs.u-bordeaux1.fr}\\
$^2$CNRS, LAB, UMR 5804, F-33270, Floirac, France \\
$^3$Universit\'e' Grenoble Alpes, IPAG, 38000 Grenoble, France \\
$^4$CNRS, IPAG, 38000 Grenoble, France \\
$^5$IRAM, 300 rue de la piscine, F-38046 Saint Martin d'H\`eres, France
}

\pubyear{2015}
\volume{314}  
\pagerange{119--126}
\jname{Young Stars \& Planets Near the Sun}
\editors{J. H. Kastner, B. Stelzer, \& S. A. Metchev, eds.}
\begin{document}

\maketitle

\begin{abstract}
Debris disks are usually thought to be gas-poor, the gas being dissipated by accretion or evaporation during the protoplanetary phase. HD141569A is a 5 Myr old star harboring a famous debris disk, with multiple rings and spiral features. I present here the first PdBI maps of the $^{12}$CO(2-1), $^{13}$CO(2-1) gas and dust emission at 1.3 mm in this disk. The analysis reveals there is still a large amount of (primordial) gas extending out to 250~au, i.e. inside the rings observed in scattered light. HD141569A is thus a ‘hybrid’ disk with a huge debris component, where dust has evolved and is produced by collisions, with a large remnant reservoir of gas. 
\end{abstract}


\section*{A debris disk still containing gas}

\begin{table}
	\centering
  \caption{Best fit parameters from DiskFit gas modeling}
 {\scriptsize
  \begin{tabular}{|c|c|c|c|c|c|c|}\hline 

      & {\bf Inclination ($^{\circ}$)} & {\bf Postion Angle ($^{\circ}$)} &  {\bf R$_{out}$ (au)} &  {\bf R$_{in}$ (au)} & {\bf T$_0$ (K)} & {\bf q} \\ 
  \hline
    $^{12}CO$ & 54.4 $\pm$ 0.4 & 86.1 $\pm$ 0.2 & 254 $\pm$ 3 & 22 $\pm$ 1 & 44 $\pm$ 2 & 0.35 $\pm$ 0.05 \\
\hline
    $^{13}CO$ & 57 $\pm$ 2 & 88 $\pm$ 1 & 253 $\pm$ 15 & 21 $\pm$ 7 & 16 $\pm$ 4 & 0.2 $\pm$ 0.3 \\
\hline

  \end{tabular}
  }
\vspace{1mm}
\end{table}

\begin{figure}[b]
\centering
\includegraphics[width=4.6in]{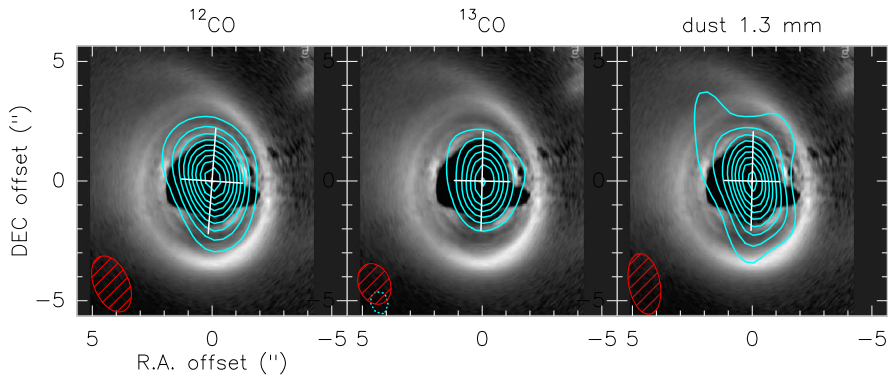} 
 \caption{Integrated intensity of the CO and dust emission at 1.3 mm, superimposed to the $HST$ scattered emission (Clampin et al. 2003). The cross indicates the position angle and aspect ratio as determined from gas modeling. Left: $^{12}$CO J=2-1 emission, contour spacing: 6$\sigma$, i.e. 5.5$\times$10$^{-1}$ Jy/beam.km.s$^{-1}$. Beam size: 2.48$\times$1.45''. Middle: $^{13}$CO J=2-1 emission, contour spacing: 3$\sigma$, i.e. 7.8$\times$10$^{-2}$ Jy/beam.km.s$^{-1}$. Beam size: 1.76$\times$1.32''. Right: continuum emission, contour spacing: 3$\sigma$, i.e. 2.0$\times$10$^{-1}$ mJy/beam. Beam size: 2.57$\times$1.31''.}
   \label{fig:maps}
\end{figure}

HD141569A is a 5$\pm$3 Myr old star (Mer\'in et al. 2004), of spectral type B9.5V/A0Ve, located 116$\pm$8 pc away (van Leeuwen 2007).
With a stellar mass of 2 M$_{\odot}$, the HD141569A system appears to be in an intermediate evolutionary stage between protoplanetary and debris disks.

A debris disk has been discovered first by $IRAS$, with an infrared excess of the same order of magnitude as $\beta$ Pictoris (L$_{disk}$/L$_{\star}$=8$\times$10$^{-3}$; Sylvester et al. 1996). 
Optical images reveal that the debris disk around HD141569A is very complex, with a double-ring architecture, a large inner depletion within 125~au, and arc and spiral features (e.g. Augereau et al. 1999, Biller et al. 2015). 
The dust appears to be of second generation origin, i.e. produced by collisions, as indicated by the timescale for collisions of $\sim$10$^{4}$ years which is $100$ times less than the age of the star (Boccaletti et al. 2003). 

In addition to its impressive debris disk, CO gas has been detected around HD141569A (Zuckerman et al. 1995; Dent et al. 2005). NIR CO and other atomic lines have also been observed (Goto et al. 2006; Thi et al. 2014). The inferred total remnant mass of gas has thus been estimated in the range 80-135 M$_{\oplus}$ (Jonkheid et al. 2006).

We present here the first resolved maps of the $^{12}$CO J=2-1 and $^{13}$CO J=2-1 emission lines, which we obtained in 2014/2015 with the Plateau de Bure Interferometer array. The integrated intensity maps of the gas are displayed in Fig.1, as well as the continuum emission at 1.3 mm.
We have modeled the data in the uv-plane using the code DiskFit (Pi\'etu et al. 2007), based on a power-law description of the physical parameters, e.g. T(r)=T$_{0}$(r/R$_{0}$)$^{-q}$.
Table 1 shows the parameters determined from this modeling for the $^{12}$CO and $^{13}$CO. The disk extends from $\sim$20~au to 250~au. From the $^{12}$CO/$^{13}$CO line ratio, the $^{12}$CO appears to be still optically thick while the $^{13}$CO is optically thin. The temperature is thus best determined from the $^{12}$CO modeling ($\sim$45~K at 100~au, a typical value for an A star). The $^{13}$CO better probes the surface density, which is here $\sim$30 times less than around typical HAeBe disks, like MWC480.  

HD141569A is thus a `hybrid' disk with a large gas component, likely primordial, and an impressive evolved debris disk. The flux at 1.3 mm is 3.5$\pm$0.1 mJy, a low value in agreement with fast evolution of the dust. The links between gas and dust properties in this and other star/disk systems have to be studied in more detail, in particular to better understand the disk dissipation/evolution mechanisms which influence the shaping of young planetary systems.

\section*{References}
\noindent Augereau, J.-C.; Lagrange, A. M.; Mouillet, D. et al., 1999, A\&A, 350,51\\
Biller, B. A.; Liu, M. C.; Rice, K. et al., 2015, MNRAS, 450, 4446\\
Boccaletti, A.; Augereau, J.-C.; Marchis, F. et al. 2003, ApJ, 585, 494\\
Dent, W. R. F.; Greaves, J. S.; Mannings, V. et al. 2005, MNRAS,277,25\\
Goto, M.; Usuda, T.; Dullemond, C. P. et al., 2006, ApJ, 652, 758\\
Jonkheid, B.; Kamp, I.; Augereau, J.-C. et al., 2006, A\&A, 453,163\\
Mer\'in, B.; Montesinos, B.; Eiroa, C. et al, 2004, A\&A, 419, 301\\
Pi\'etu, V.; Dutrey, A.; Guilloteau, S., 2007, A\&A, 467, 163 \\
Sylvester, R. J.; Skinner, C. J.; Barlow, M. J. et al., 1996, MNRAS, 279, 915\\
Thi, W.-F.; Pinte, C.; Pantin, E. et al., 2014, A\&A, 561, 50\\
van Leeuwen, F., 2007, A\&A, 474, 653\\
Zuckerman, B.; Forveille, T.; Kastner, J. H, 1995, Nature, 373, 494\\

\end{document}